# Spin-Based Neuron Model with Domain Wall Magnets as Synapse

Mrigank Sharad, Charles Augustine, Georgios Panagopoulos and Kaushik Roy, *Fellow, IEEE*

*Abstract*—We present artificial neural network design using spin devices that achieves ultra low voltage operation, low power consumption, high speed, and high integration density. We employ spin torque switched *nano-magnets* for modelling neuron and *domain wall magnets* for compact, programmable synapses. The spin based neuron-synapse units operate locally at ultra low supply voltage of 30mV resulting in low computation power. CMOS based inter-neuron communication is employed to realize network-level functionality. We corroborate circuit operation with physics based models developed for the spin devices. Simulation results for character recognition as a benchmark application shows 95% lower power consumption as compared to 45nm CMOS design.

## I. INTRODUCTION

Hardware implementation of computation architectures based on artificial neural network (ANN) has always been challenging in terms of power consumption, level of integration and throughput. Prior work in this field involved development of circuit models for neurons and synapses using CMOS [1-5]. Digital ANN designs proposed earlier, consume large area and hence limit the level of integration [31]. On the other hand, analog designs, although compact, consume large amount of power [2].

In order to tap the potential of neural network based computation at the hardware level, the device-circuit models for the neuron and the synapse, apart from being compact, should also achieve low power consumption. In this work we propose the application of spin-devices in ANN hardware design that can help achieve these goals.

Recent experiments on lateral spin valves (LSV) have shown spin-torque induced switching of *nano-magnets* using spin-polarized current flow through metal channels [7, 8]. Such magneto-metallic LSV's can operate at ultra-low terminal voltages, resulting in low switching energy [10, 11]. A multi-input LSV can perform non-Boolean, analog-mode computation like majority-evaluation [9].

M. Sharad, and G. Panagopoulos and K. Roy are with the School of electrical and Computer Engineering, Purdue University,West Lafayette, IN 47907 USA. C. Augustine is with the Circuit Research Lab, Intel Corporation, Hillsboro, OR, US A.

All spin logic (ASL) design based on majority evaluation using spin torque in LSV's has been proposed previously [10]-[16]. We show that, with an appropriate clocking scheme, a spin majority gate with weighted inputs mimics the neuron-synapse functionality. Programmable spin injection strength of domain wall magnet can be used to implement a compact synapse. In the proposed neuron-synapse model, charge current flows through a low resistance path that constitutes of the *nano-magnets* and non-magnetic metal channels. This allows application of ultra low terminal voltages, resulting in low power consumption.

Energy dissipation for spin mode computation increases steeply with the separation between *nano-magnets*. This is due to the limited spin diffusion length of non-magnetic channels [10, 11]. Hence, spin-mode signaling between two neuron units proves inefficient. Therefore, we employ CMOS based, charge-mode inter-neuron signaling scheme in order to realize network-level functionality. Hence, the programmable, spin-CMOS hybrid ANN architecture, presented in this work, clubs the benefits of localized, spin based, low-energy computation and robust charge-mode communication.

The rest of the paper is organized as follows. Section.2 describes the operation of spin majority-gate based on lateral spin valve (LSV). Detail description of the proposed neuron-synapse model is given in section.3. Section.4 discusses system level integration. Device simulation framework employed in this work is discussed in section 5. Performance of the spin based ANN design for a benchmark application (character recognition), and its comparison with 45nm CMOS analog and digital designs is given in section 6. Summary and conclusions are given in section 7.

## II. MAJORITY GATE BASED ON LATERAL SPIN VALVE

Two different methods of current induced STT based switching of *nano-magnets* have been proposed in recent years. The first involves injection of spin polarized charge current into a *nano-magnet*. The second strategy, on the other hand, employs pure spin-current injection for flipping a *nano-magnet* [7, 8]. Fig. 1a and Fig. 1b shows the lateral spin valve (LSV) structure with local and non-local spin current injection respectively [7, 8]. It consists of an injecting magnet and a receiving magnet connected through a non-magnetic channel. Electrons flowing into the channel through the transmitting magnet (which possesses 'up-spin' polarization) get up-spin polarized when they reach the magnet-channel interface. Spin-polarized charge-current is modeled as a four-component quantity, one charge component and three spin components $(Is_x, Is_y, Is_z)$ [11, 12]. For the non-local case, the charge component of the input current flows into the ground lead. The output magnet-channel interface absorbs the transverse spin components of the current which in turn exerts spin torque on the output magnet and causes it to flip. Owing to the separation of the spin diffusion current responsible for *nano-magnet* switching, from the charge current flow, spin transport in the lateral spin valve is often termed as 'non-local'. On the other hand, in the case of local spin injection, the spin polarized charge current input through the first magnet is injected into the output magnet.

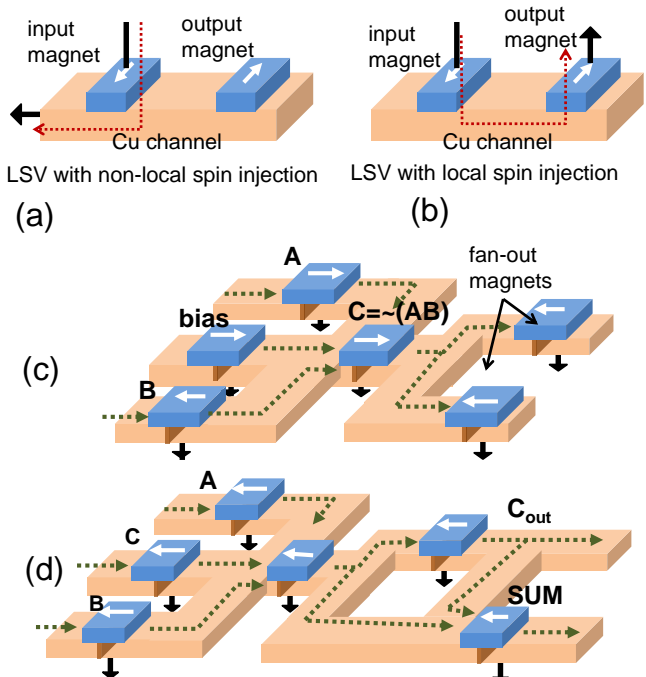

Figure 1(a) Lateral spin valve with non-local spin injection (b) LSV with local spin injection (b) ASL spin majority gate for NAND logic [10], (c) compact ASL Full adder using five magnets [13].

Analog characteristics of current mode switching employed in LSV's can facilitate non-Boolean computation like majority evaluation. Hence, LSV's with multiple input magnets can be used to design spin majority gates [9]. In [10] authors proposed 'all spin logic' (ASL) scheme that employed cascaded LSV's interacting through unidirectional, non-local spin current [12]. Fig 1c and fig. 1d depict ASL NAND gate [10], and, ASL full-adder [13], based on spin-majority evaluation.

A clock synchronized operation of the spin majority gate with fixed input magnets can be compared to that of a neuron, if the output magnet's state is restored after every flipping. The two spin-polarization states of the input magnets are analogous to bipolar, binary synapse weights with values +/-1. In this work we propose the use of domain wall magnets as input synapse to realize programmable, bipolar, multi-level weights for a spin-based neuron model. To reduce the amount of average current injection per synapse we incorporate current mode Bennett clocking in the neuron model [10]. It involves switching the *nano-magnet* to an intermediate meta-stable state from which, it can be switched back to one of its stable states with a very small current. In this work output magnet of the proposed neuron model is switched with non-local spin torque, i.e. with pure-spin current. However, in [34] we showed that a device with local spin injection can also be used in the proposed design scheme.

As mentioned earlier, due to limited spin diffusion length of metal channels, spin-mode signaling between neurons can be inefficient. Moreover, physical layout of random interconnects between multiple neurons using planar LSV structure becomes challenging. Hence, we employ CMOS based charge-mode signaling for long-distance inter-neuron communication. The proposed design therefore, exploits ultra low voltage operation of spin neurons along with robust charge mode signaling to realize network functionality.

### III. SPIN BASED NEURON-SYNAPSE MODEL

In this section we present the spin based neuron-synapse model. First we discuss the application of domain wall magnet as a synapse. Following this, the neuron model is described which is based on the lateral spin valve structure discussed in Section-2.

#### A. Domain wall magnet as synapse

*Domain wall magnet* (DWM), shown in fig. 2a, consists of two ferromagnetic domains separated by a non-magnetic region or *domain wall* (DW). Domain wall is formed in a magnetic nano-strip due to balance in anisotropy and exchange energies present in *nano-magnet* [18]. Domain wall can be moved along a magnetic nano-strip by application of magnetic field [18] or by injection of charge current along the nano-strip. [19]. Fig. 2b shows the simulation plot for domain wall velocity vs. injected current density, benchmarked with experimental data in [20].

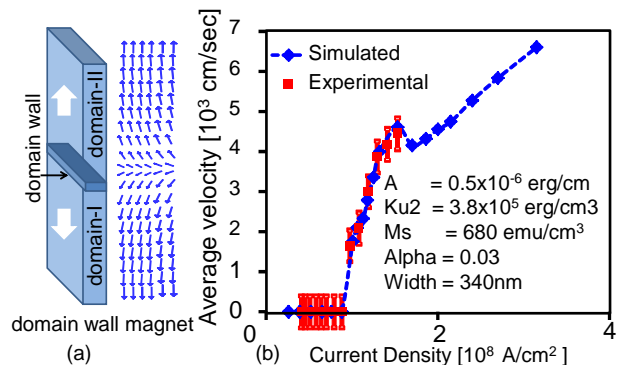

Fig. 2(a) Fig. 2.Domain wall magnet (b) DW velocity as a function of current density with experimental data in [20].

Application of DWM in the design of non-volatile memory [21] and logic design [22] has been explored by several authors. In the present work, we propose the use of DWM as synapse, where its programmable spin injection strength is used for implementing spin-mode weighting operation. Fig. 3a shows a domain wall magnet interfaced with the non-magnetic channel of a neuron.

In order to write the weight into the DWM, current is injected along the length of the domain wall as shown in fig. 3a. Under this condition the channel is kept in a floating state. A thin MgO layer incorporated at the top and bottom surface of the DWM reduces the fringe current passing through the parallel path provided by the floating channel and the input lead, during the write operation. The interface oxide also imparts an effective resistance to the input lead of the DWM that makes it dominate the parasitic resistance of the signal-routing metal-lines.

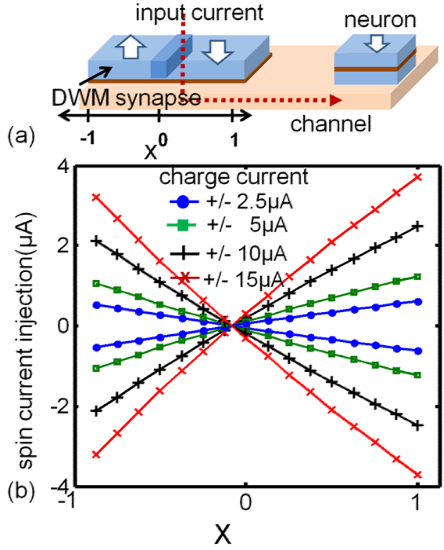

Fig. 3 (a) Domain wall synapse with channel interface (b) Spin polarization strength current injected through DWM as a function of DW location

During computation, the input current is injected into the channel through the domain wall in the vertical direction. Fig. 3b shows the plot for spin polarization of current passing into the channel through the DWM vs. domain wall location for different charge current values. It can be observed that, spin polarization strength of the charge current reaching the channel is proportional to the offset of the domain wall location from the centre. For the extreme left location of the domain wall, the charge current reaching the metal channel is maximally up-spin polarized and vice-versa. The net polarization is reduced to zero for the central location of the domain wall, as equal amount of up and down spin electrons are injected into the channel in this case.

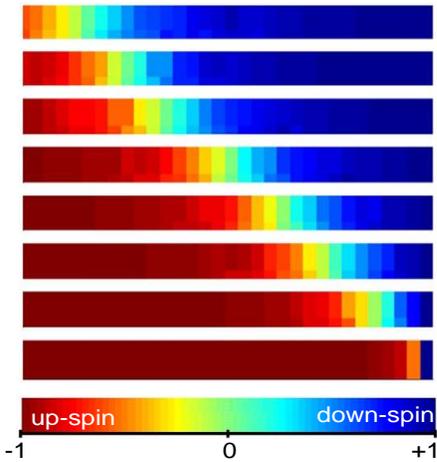

Fig. 4 Magnetization state of the DWM at equal time intervals after starting of DWM motion.

In the simplest case, the two extreme locations of the domain wall can be employed for implementing programmable binary weights. Neural networks with binary weights can be applied for logic synthesis and pattern recognition applications [28, 29]. However, network with binary weight may require larger number of neurons for a given operation, as compared to a network with higher number of weight levels depending upon the size of the exhaustive training set [29]. Larger number of weight levels can be obtained by employing longer DWM stripes that can facilitate better quantization of domain wall location. It has been shown that incorporation of nano-scale notches in the DWM strips can enhance the stability of DW at the notch sites [23]. The incorporation of notches along the length of the DWM synapse can help in achieving higher writing accuracy. In this work we incorporate DWM synapses with a cross section area of 350x80nm$^2$. Notches etched at 22nm interval along the 350nm long DWM strip can provide 16 levels of weight. Fig. 4 shows the magnetization state of the DWM at equal time intervals after the application of 250psec voltage pulse train.

Physics based device modeling of domain wall synapse is discussed in section 5

B. *Spin based neuron model*

Transfer function of an 'integrate' and 'fire' neuron is given by eq. 1.

$$Y = f\left(\sum w_i I_i + b\right) \quad (1)$$

Here, $w_i$ and $I_i$ are the weights and corresponding inputs and $b$ is the neuron bias. The bias can be chosen to be zero. It however aids in training convergence and can be easily implemented by an additional synapse magnet which is driven by a clock. The function $f(x)$ is given by eq.2 and approximates a step transfer function for a sufficiently large $N$.

$$f(x) = \left(1 + e^{-N(x-t_o)}\right)^{-1} \quad (2)$$

Here $t$ denotes the threshold of the neuron. It can be inferred that a higher $|t|$ would require a larger value of $|x|$ to switch the neuron. For a given set of normalized weights $W_i$, this translates to larger levels of the input signals $I_i$. For the spin based neuron model, this implies larger input current per synapse and hence higher power consumption. Therefore, switching threshold of the output *nano-magnet* needs to be reduced. We incorporate current-mode Bennett-clocking to achieve this.

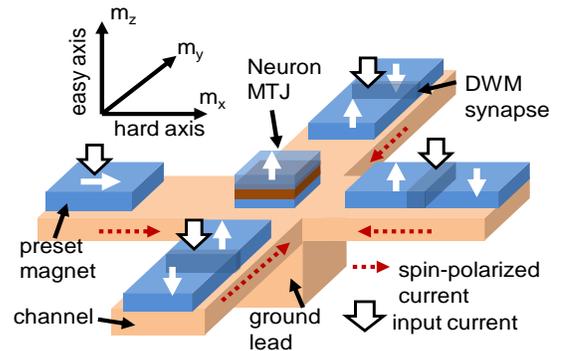

Fig. 5 Spin-based neuron-model with three inputs (DWM synapses). The free layer of the neuron-MTJ is in contact with the channel and its polarity, after preset, is determined by spin polarity of combined input-current in the channel region just below it.

The device structure for the neuron with three inputs is shown in fig. 5. The 'firing-magnet' forms the free layer of an MTJ. The two anti-parallel, stable polarization states of a magnet lie along its easy axis (fig. 5). The direction orthogonal to the easy axis is an unstable polarization state for the magnet and is referred as its hard axis [10, 13]. The preset-magnet shown in fig. 5 has its easy axis orthogonal to that of the neuron magnet (MTJ free-layer which is in contact with the channel). In the beginning of a clock-period, current-pulse injected through the preset-magnet forces the neuron-magnet to the hard-axis configuration (fig. 6). As soon as the hard-axis biasing-pulse goes low, the free-layer makes transition to the easy-axis polarity governed by the polarity of net spin-polarization of the channel-current flowing under it. As a result, the firing-magnet, i.e., the free layer of the MTJ acquires either parallel or anti-parallel polarization with respect to the fixed-layer. Note that, summation of the 'spin-weighted' input currents (eq. 1), received through multiple DWM synapses, takes place in the metal-channel. Whereas, the symmetric step-transfer function upon the summed spin-current (eq. 2), is realized with the help of Bennett-clocking of the neuron-magnet.

When the clock is low, a CMOS-based detection unit (discussed later) reads the state of the neuron MTJ. For a parallel configuration, it generates a high output whereas for the anti-parallel configuration, it settles to a low value. Hence, the detection unit converts the spin-mode information of the neuron magnet's state into a charge-mode signal. For a particular stage of network, spin and charge mode evaluations occur in alternate clock phases (fig. 6). For a multistage, feed-forward neural network, neurons in alternate stages are driven by complementary clock phases. This results in a fully parallel and pipelined network.

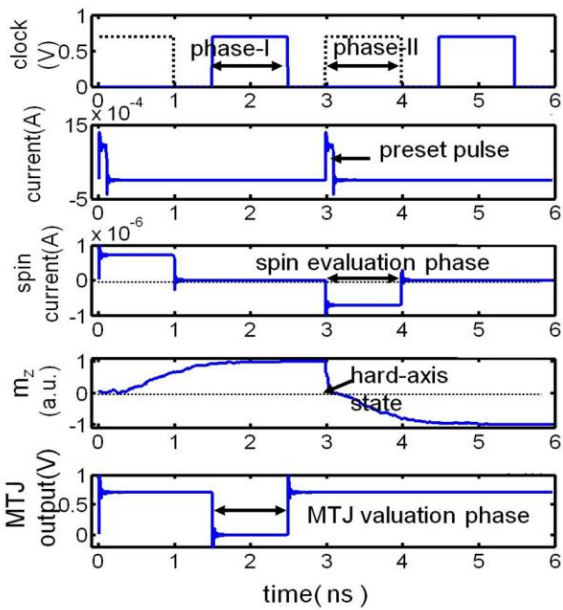

Fig. 6 Timing waveform for the proposed neuron model

In the proposed neuron model, the use of non-local STT switching allows a low resistance path for static charge current flow that includes the DWM synapse and the non-magnetic channel. This allows application of very small voltages, which in turn results in ultra low energy operation for the magneto-metallic neuron-synapse unit. The detection scheme, discussed later, involves negligibly small transient current flow through the high resistance MTJ stack.

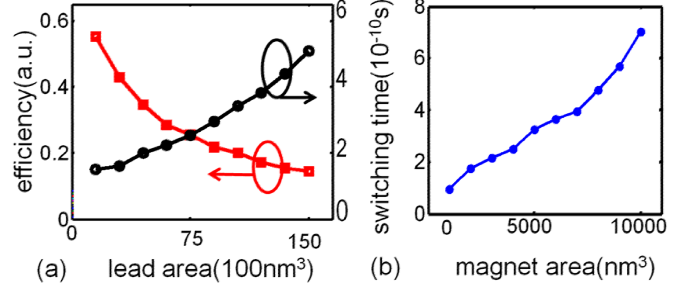

Fig.7 (a) Increase in spin injection efficiency and switching speed through scaling of ground lead for a fixed current input (b) Reduction in switching time with combined scaling of neuron magnet for a fixed current input.

Performance metrics of the neuron-device, like, spin injection efficiency, switching energy and switching-speed can be improved by the appropriate choice of magnet parameters, device geometry and operating conditions.

Non-local spin injection efficiency in the device can be defined as the ratio of spin current $I_s$, injected into the output magnet and the net spin polarized charge current in the channel under the neuron MTJ. As discussed earlier, the spin components of the combined synapse current gets divided between the output magnet and the ground lead. Thus the spin injection efficiency for a given charge current input is enhanced by increasing the resistance of the ground lead (fig. 7a).

Smaller volume for the output magnet, along higher coercive field $H_k$ leads to higher switching speed for a given spin current (fig. 7b) [11]. It also leads to faster easy axis restoration (fig. 8a). In order to maintain the spin injection efficiency, resistance of the ground lead needs to be scaled up proportionately.

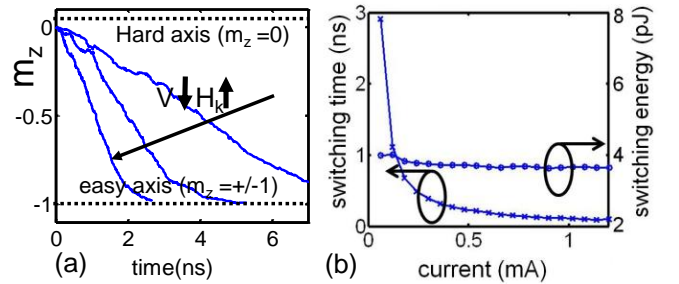

Fig. 8(a) Increase in easy-axis restoration speed with $H_k$ and reducing magnet volume (for spin current of 0.5 μA) (b) Hard-axis switching time and switching energy vs. switching current.

Hard-axis switching-energy is a significant portion of the energy dissipation per-neuron, per-cycle. Fig. 8b shows that, the hard-axis switching current increases with switching speed (~direct proportionality [11]). Hence for a given terminal-voltage, the switching-energy remains almost constant. In the present work, the hard axis biasing current is supplied through a transistor operating between a small terminal voltage. In order to allow a small transistor width and hence, lower clocking power, it is favorable to choose the smallest possible value for switching current and hence maximum possible preset pulse width for a given operating frequency. In this work we employed preset-current pulse of amplitude 300μA and pulse width 0.5ns.

### C. Modular neuron-synapse unit

A centre-surround layout for a neuron with 12 input synapses is shown in fig. 9. Spin-polarized charge current inputs from DWM synapses combine in the channel and flow into the ground lead located near the neuron MTJ. Spin polarization strength of charge current decays exponentially with the distance travelled along the non-magnetic channel. Thus, the channel-length between the synapses and the neuron must be within 1-2 times spin flip length ($\lambda$) [10, 11]. This imposes a limit on the number of input synapses for the structure shown in fig. 9. For copper channel ($\lambda\sim 1\mu m$) up to ~32 synapses can be combined directly. For graphene channel ($\lambda\sim 6\mu m$) this number can be increased.

Limited spin-diffusion length also introduces mismatch between the strengths of different DWM synapses, depending upon their location with respect to the neuron magnet. The two synapses $S_1$ and $S_2$ depicted in fig.9 are the closest and the farthest synapse from the neuron magnet respectively. For a neuron with 16 input synapses this effect does not introduce a significant mismatch (fig. 10a). However, for a 32 input neuron, the mismatch is quite prominent (fig. 10b). The mismatch can be mitigated by slightly grading the magnitude of synapse current injection into the DWM synapses so as to equalize all the weights (fig. 10b).

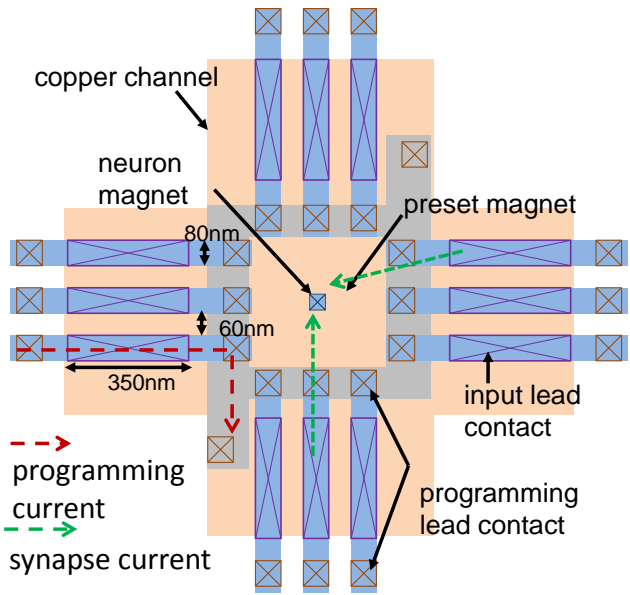

Fig.9.Centre-surround layout of the proposed neuron-synapse unit. Spin-weighted current inputs from DWM synapses combine in the central region of the 2-D metal channel, where the neuron is located

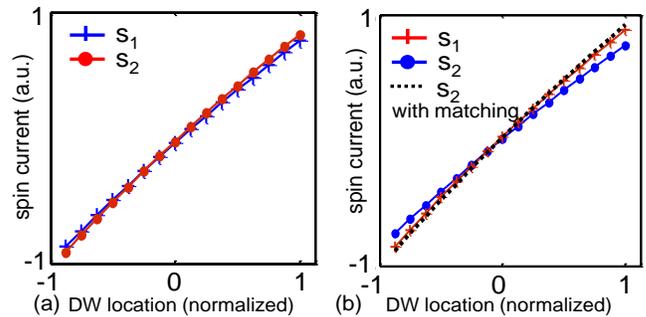

Fig. 10 (a)Location dependent synapse weight mismatch for 16 input neuron (b) Mitigation of synapse weight mismatch for 32 input neuron through enhanced current injection into weaker synapse.

Domain wall programming interface is also depicted in fig. 9, where the contact-via's and path for DWM writing current flow have been indicated. Selection of a pair of transmitting and receiving neurons indentifies the synapse to be programmed. Thus, only two transistors per neuron, (for identifying it as receiving or transmitting neuron) suffice for programming the whole network. In the case of a cellular architectures based on arrays of identical neuron units, the whole array can be programmed parallelly [31-32].

Figure. 11 depicts the plot for spin potential in the central-region of the channel, surrounding the output magnet of a 16 input neuron, under firing and non-firing conditions. It shows that, in case of a firing event, the entire channel is dominantly at a positive spin-potential and vice-versa.

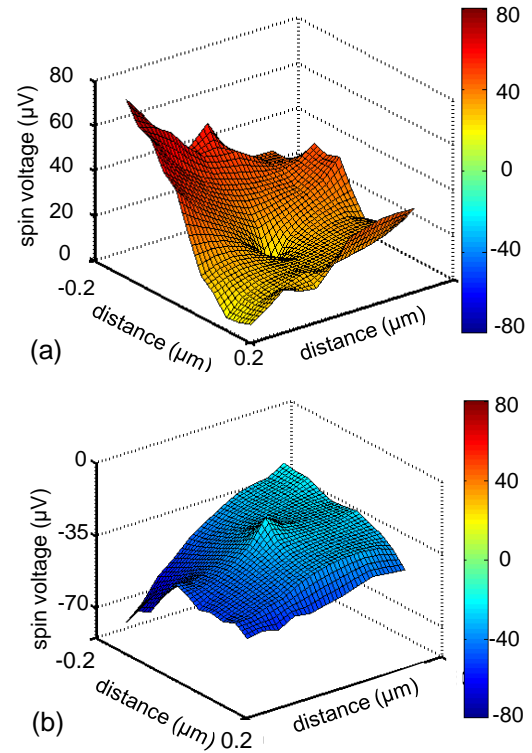

Fig. 11 (a) Channel spin potential of a 16 input neuron under firing condition (b) Channel spin potential under non-firing conidtion

## IV. SYSTEM INTEGRATION

Due to small spin diffusion length of metal channels, spin-mode signaling cannot be used for network connectivity. Hence, in this work the spin-based neuron-synapse modules are interconnected through charge-mode signaling using CMOS. The spin-mode 'firing' information is converted into charge-mode signal using the dynamic CMOS latch, shown in fig. 12a.

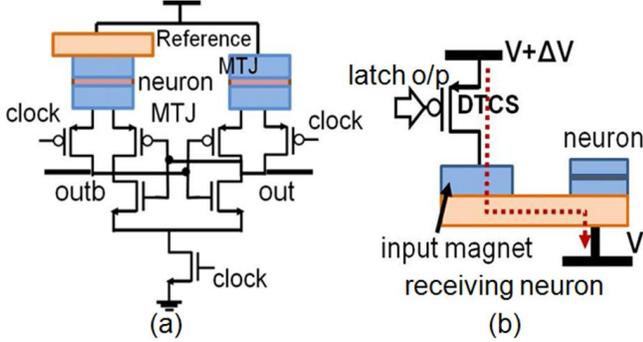

Fig. 12 (a) Differential MTJ latch (b) Inter-neuron current-mode signaling using deep triode current source (DTCS) transistor.

It compares the effective resistance of the MTJ units in its two load branches. The firing MTJ of the neuron unit connects to one of the loads, whereas, a reference MTJ is connected to the other.

The latch drives a distributed set of current source transistors which in turn supply charge current to all receiving neurons through the respective input magnets (DWM) (fig. 12b). The source terminal of the current source transistors and the ground lead of the spin based neuron modules are biased at $V+\Delta V$ and $V$ volts respectively. Hence, the synapse current flows across a small terminal voltage of $\Delta V$. In the present work, values of $V$ and $\Delta V$ are chosen to be 800mV and 30mV respectively. The CMOS units operate between 800mV and 0V. Biasing of the spin modules between two relatively high DC levels proves advantageous as compared to direct application of a small supply voltage of magnitude $\Delta V$. This is because, application of differential DC supply can mitigate the impact of I-R voltage drop along the supply lines. It can also be exploited to reject the common-mode noise in the dual supply lines. Moreover, generation of clean DC levels below 100mV is challenging in the state of art CMOS technology, whereas a regulated voltage supply of higher magnitude can be distributed with less than 0.1% fluctuation [30].

For supplying a current of 5µA per synapse (across a drain to source voltage of 30mV) for 16 receiving neurons, the required source transistor width in 45nm technology is around 2.5µm. In order to minimize the impact of synapse current mismatch, distributed source transistors are used.

Fig. 13 depicts the correspondence between the proposed spin-CMOS hybrid ANN and the biological neural network. The spin potential of the 2-D metal-channel (which is analogous to neuron cell body) depicted in fig. 11, can be related to the electrochemical potential in biological-neuron's cell-body [33]. Inter-neuron communication in the present design is realized using ultra-low voltage current transmission, which is somewhat similar to the natural mechanism [33]. However, the aim of the proposed model is not to mimic the biological neural network in terms of functionality, but to evolve a model for artificial-neural-network suitable for computational hardware.

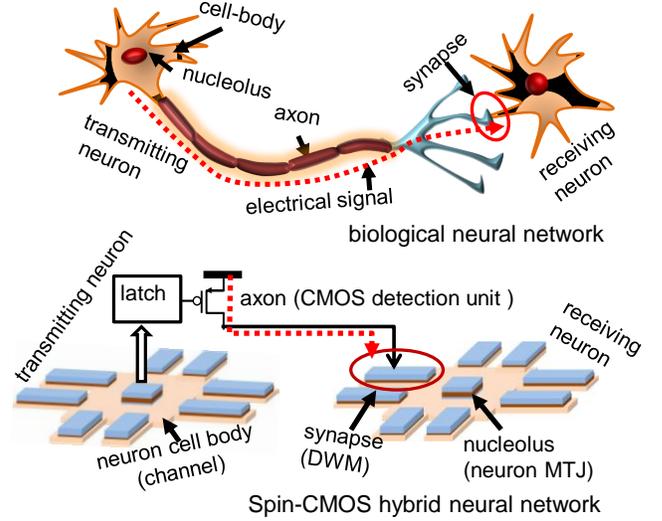

Fig. 13 Correspondence of the spin-CMOS Hybrid ANN to biological neural network

## V. SIMULATION FRAMEWORK

In this section we describe the physics based simulation framework used in this work for simulating the spin-based neuron-synapse units.

In order to simulate the neuron model, which is based on the lateral spin valve structure shown in fig. 1a, we self-consistently solve both the transport and the magnet dynamics equations using the four component spin-circuit model [12]. In this simulation framework, the channel spin transport is based on the spin diffusion model developed by Valet–Fert [26], The magnet-channel interface is modeled based on the interface model developed by Brataas *et al.* [27]. Both these models are well established and are used for spin transport in long channels [11, 12]. The spin diffusion formulation yields four component conductance matrices $G_{magnet}$, $Gl_{ead}$, $G_{int}$ and $G_{ch}$ for the elements of nano-magnets, supply leads, magnet-channel interface and the non-magnetic channel, respectively. The four components are the charge and the three spin components. The conductance matrices relate four component voltage drop and current flow between different circuit nodes,

$$[I_c, I_c^z, I_c^x, I_c^y] = [G]_{4\times 4} [V_c, V_c^z, V_c^x, V_c^y] \quad (3)$$

The non-magnetic channel and lead elements are modeled as π-conductance matrices with shunt $G_{sh}$ and $G_{se}$ as shunt and series components, respectively [11].

$$G_{sh} = \begin{pmatrix} 0 & 0 & 0 & 0 \\ 0 & g_{sh} & 0 & 0 \\ 0 & 0 & g_{sh} & 0 \\ 0 & 0 & 0 & g_{sh} \end{pmatrix} \quad G_{se} = \begin{pmatrix} \frac{A}{\rho l} & 0 & 0 & 0 \\ 0 & g_{se} & 0 & 0 \\ 0 & 0 & g_{se} & 0 \\ 0 & 0 & 0 & g_{se} \end{pmatrix} \quad (4)$$

Here, $g_{sh} = (A/\rho\lambda)\tanh(l/2\lambda)$ and $g_{se}=(A/\rho\lambda)\mathrm{csch}(l/\lambda)$, l is the length of the contact, A is the area of the contact, ρ is the resistivity and λ is the spin-flip length. These conductance matrices are obtained by solving spin-diffusion equation as shown in [11]. Contact-magnet-channel interface can be described through the matrix $G_{int}$.

$$G_{int} = \begin{pmatrix} g & gP & 0 & 0 \\ gP & g_{se} & 0 & 0 \\ 0 & 0 & \Gamma+\Gamma^* & i(\Gamma-\Gamma^*) \\ 0 & 0 & -i(\Gamma-\Gamma^*) & \Gamma+\Gamma^* \end{pmatrix} \quad (6)$$

where, $g=2-r_l r_l^* - r_r r_r^*$ and $gP=r_r r_r^* - r_l r_l^*$, $\Gamma=1-r_l r_r^*$ and P is the polarization of magnet. $r_l$ and $r_r$ are the reflection coefficients correspond to left and right spin, respectively. The components of the interface matrix are dependent upon the *nano-magnet's* magnetization state, to be evaluated self consistently with magnet dynamics. Note that the elements of $G_{sh}$ are responsible for the decay of spin current along the channel due to spin diffuse scattering [11].

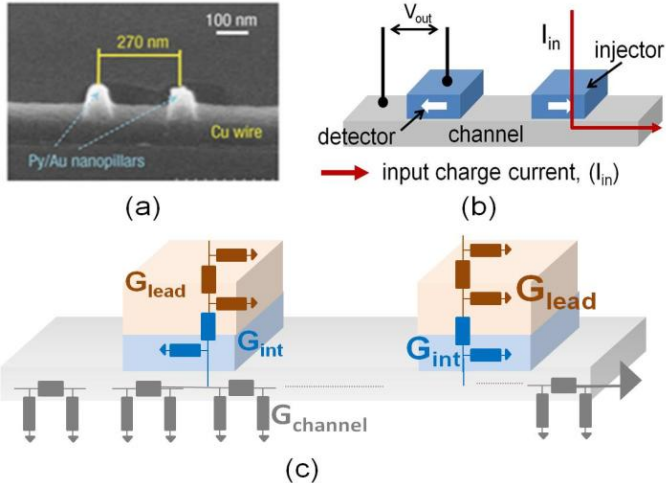

Fig. 14 (a) Fabricated LSV structure in [7], (b) Depiction of structure in fig.1 a, (c) Spin circuit model based on spin diffusion model for the device in fig. 1a

The *Nano-magnet* dynamics is captured by solving the Landau-Lifshitz-Gilbert equation (eq. 7), self-consistently with spin diffusion.

$$\frac{d\hat{m}}{dt} = -|\gamma|\hat{m}\times\vec{H} + \alpha\hat{m}\times\frac{d\hat{m}}{dt} - \frac{1}{qN_s}\hat{m}\times(\hat{m}\times\vec{I}_s) \quad (7)$$

Here m is the magnetization vector, α is the damping constant, $N_S$ is the number of spins in the magnet, γ is gyromagnetic ratio, H is the effective magnetic field and $I_S$ is the spin-current, which is obtained by the transport framework. This simulation-framework has been benchmarked with experimental data on LSV's [10-12]. This approach leads to the mapping of a spin device structure, involving *nano-magnets* interacting through non-local spin transport, into an equivalent "spin-circuit" [10]. The circuit model for the lateral spin valve is shown in fig. 14. The spin circuit approach, discussed above, is extended to the 2-D neuron-synapse model shown in fig. 9, where the channel is modeled as a 2-D grid of 10nm x 10nm sections.

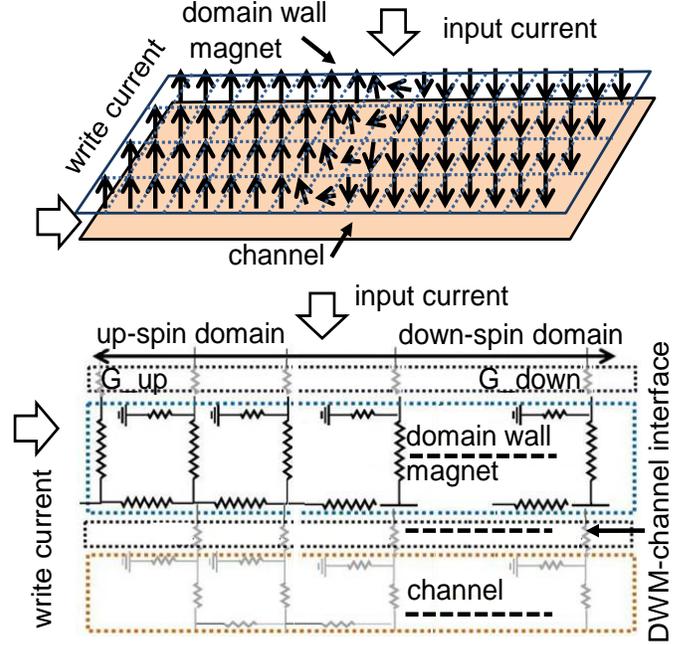

Fig. 15. Domain-wall simulation model

The device model for the domain wall structure is derived from the aforementioned spin diffusion model. It consists of a 2-D grid of *nano-magnets* obtained by dividing the nanostrip into square grids (10nm x 10nm) as depicted in fig. 15. Each *nano-magnet* is modeled as a ∏ conductance-network with shunt and series components $G_{0F}$ and $G_F$ (Four Component Spin Transport model), respectively, using Valet-Fert diffusion model [26] and interface model by Brataas [27]. The resulting spin circuit is shown in fig. 15. It yields the spin current components at each lattice points for a given input voltage. These spin currents are used to evaluate LLG at each point to capture the *nano-magnet* dynamics. The conductance matrices are dependent upon the magnetization state of the grid points and hence, the spin diffusion transport is solved self consistently with LLG at each grid point. We benchmarked our simulation framework for DWM with experimental data in [20]. The corresponding plot for DWM velocity as a function of charge current density is shown in fig. 2.b. The effect of channel interface on the writing process is incorporated by including the *nano-magnet*-channel interface conductance matrix in series with the channel conductance matrix at each grid point as shown in fig. 15. The interface conductance matrix constitutes of spin dependent conductance components for MgO [16].

As discussed earlier, during computation, the input current is injected into the channel through the domain wall in the vertical direction. Hence, writing and computation modes are fully decoupled. Therefore, for the computation mode, the

DWM synapses can be modeled as two parallel *nano-magnets* with opposite polarities and area dependent on the domain wall location i.e., the weight.

## VI. NETWORK SIMULATION

In this section we describe the network simulation for character recognition as a benchmark application. Impact of process variation upon network performance is assessed. We also compare the performance of the proposed spin-CMOS hybrid ANN with that of a state of art CMOS ANN design.

### A. Benchmark Application

We simulated character recognition as a benchmark application for the proposed spin-CMOS hybrid design. The overall process for character recognition can be divided into two steps, namely, edge extraction and pattern matching. For edge extraction, column wise pixels form the binary image along four directions - horizontal, vertical and $\pm 45°$ are fed to the first stage neurons.

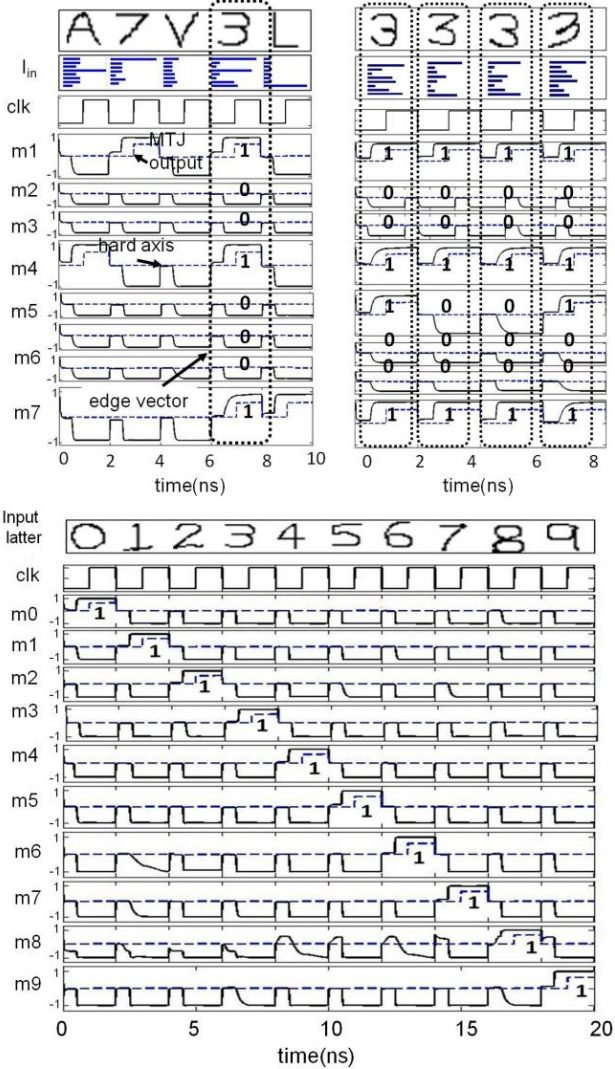

Fig. 16(a) Barcode generation for horizontal edges in alphanumeric characters, (b) Effect of style variation on horizontal bar code, (c) Output waveforms for numeric character recognition.

These neurons generate a high output if the number of non-zero pixels along a particular column (or equivalently the spin current input $I_{in}$ to the neuron) is higher than the neuron threshold. Note that, a desirable threshold for a neuron is set by applying a bias input to it. The horizontal edge extraction process for different input character is depicted in fig. 16a. The solid lines denote the magnetization state of the neuron magnets whereas the dashed lines indicate the corresponding MTJ evaluation. Fig. 16b shows the effect of variation in the handwriting style for the numeral '3' on the horizontal bar code. It shows that, significant variations in writing style translate to slight variations in the barcode pattern which can be tolerated by an ANN. Variation tolerance can be enhanced by training with different styles of input characters. The resultant four binary patterns form a 1-D representation of the input character. This pattern is fed to the output stage of the network for classification. The output neurons correspond to the 36 alpha numeric characters. The output evaluation for numeric characters is shown in fig. 16c.

### B. Variation analysis

As described earlier, variation aware circuit design techniques, like, the use of distributed and matched current source transistors, can reduce the effect of CMOS process variation upon network performance significantly. The impact of nano-magnet parameter variation upon system performance however, needs to be assessed while modeling an ANN with nano-scale devices.

The critical DWM parameters, having impact on computation accuracy, can be identified as, interface oxide thickness, cross section area and domain wall locations. Variation in oxide thickness can lead to mismatch in the effective resistance of the DWM input leads. This leads to difference in charge current injection for different synapses, which in turn introduces errors in weights. However, since the interface oxides are generally grown through atomic layer deposition (ALD), their thickness can be precisely controlled. Cross section area variation in the DWM synapse leads to variation in spin polarization of the input charge current. Inaccuracy in domain wall programming directly translates to imprecision in synapse weights.

The effect of writing inaccuracy in the domain wall synapse is captured in the simulation framework by imposing random shifts in domain wall location (fig. 17a). Impact of process variation like line-edge roughness (LER) is incorporated in terms of random variations in the DWM cross section area (fig. 17a). Fig. 17b shows the superimposed effects of inaccurate writing and geometrical imperfection upon DWM weight.

The neuron magnet is highly scaled in order to achieve fast easy axis restoration and lower switching current. It is therefore expected to be prone to thermal noise and magnet parameter variations. Fig. 18a depicts the effect of thermal noise on neuron transfer characteristics. Under very small input spin current, the easy axis restoration can be non-deterministic due to thermal noise. The impact of the noisy transition zone on overall network performance can be ignored as long as it correspond to a small fraction (<10%) of the range of spin current injection $I_s$. The range of $I_s$ in turn depends linearly on average synapse current $I_{in}$(fig. 18b). Hence, noise determines the limit to which the average synapse current can be lowered to reduce the overall power consumption.

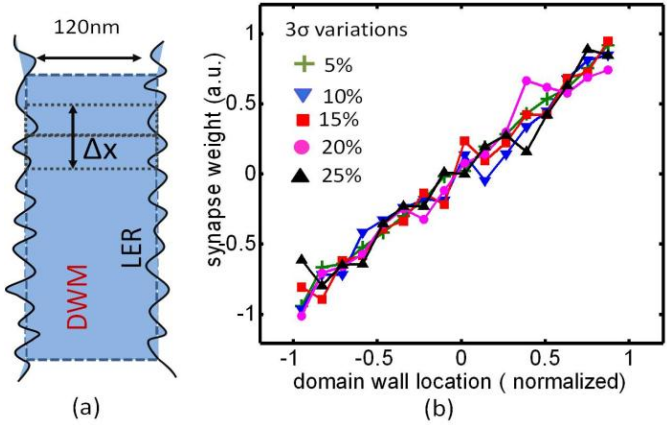

Fig. 17(a) DWM cross section area showing LER(b) Combined effect of LER, and programming inaccuracy upon DWM weight.

Since, Bennett clocking places the neuron switching threshold at origin, irrespective of the magnet parameters, the impact of output-magnet parameter variations upon the device transfer characteristics is significantly mitigated. Parameter variation however, affects the dynamic switching characteristic of the neuron. Easy axis relaxation time for neuron magnet spreads with increased parameter variations, which limits the maximum operating frequency for reliable operation. Fig. 18c shows the scatter plot for neuron switching time for two different sizes of the output magnet. The input current has been varied over two orders of magnitude (20µA to 0.5µA) corresponding to the variation in synapse currents for different input combinations. 25% 3σ variation has been applied for critical magnet parameters. It is evident that lower volume and higher $H_k$ (for a constant switching energy barrier) results in lower spread and hence, facilitates higher operating frequency.

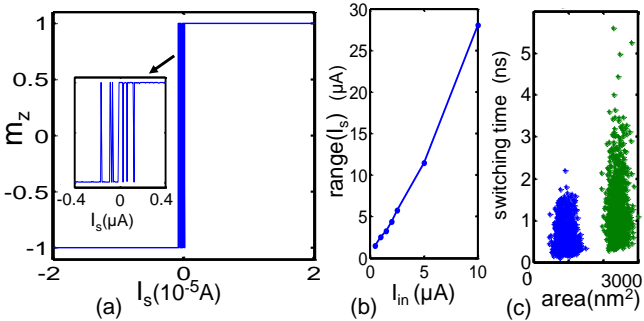

Fig. 18(a) Noisy threshold region for the neuron-magnet due to thermal-noise (b) Range of spin-current ($I_s$) injection into neuron magnet vs. synapse current($I_{in}$) for a neuron with 16 synapse (c) Scatter plot for easy axis relaxation time under parameter variation and varying input current.

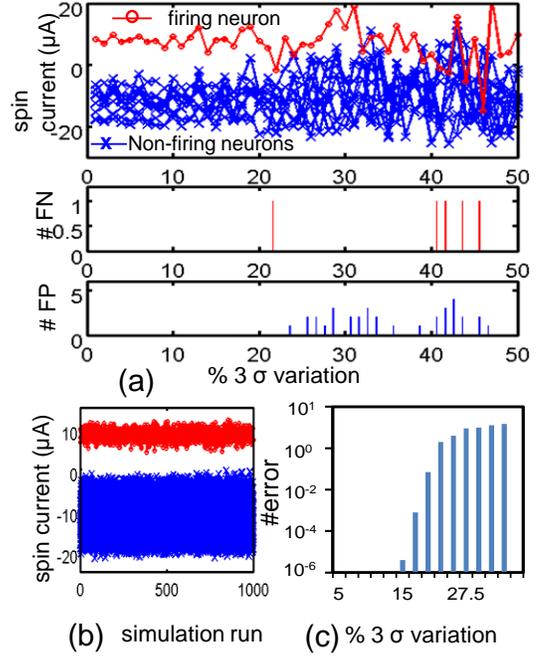

Fig. 19 (a) Impact of process variation on spin current input to neuron magnets, (b) 1000 point simulation for 15% 3σ variations, (c) Monte Carlo results for a neuron under combined process variations

Fig. 19a shows the effect of increasing process variation upon the spin-current delivered to the output neurons corresponding to the input character. A negative value of spin current for firing neuron and a positive value of spin current for a non-firing neuron denotes an error. The resulting false negatives (FN) and false positives (FP) are shown in the figure.

Network simulations show that, among different device parameters considered, domain wall location has the maximum impact upon network performance. This is because it bears a direct relation to the synapse weight. As mentioned earlier, incorporation of nano-scale notches along the DWM length can achieve improved programming accuracy. Fig. 19b shows the plot for 1000 simulation points for the network under combined 15% 3σ variations for DWM and neuron magnets. Monte Carlo simulation results for a neuron given in fig. 19 c depicts that it retains accuracy up to more than 18% 3σ variations. Note that 18% variation in a 16 level synapse weight implies a programming error of 3 levels.
.

C. *Design Performance*

In order to establish a comparison with state of art CMOS technology we implemented the same network architecture in CMOS 45nm technology in two different ways, digital and analog. For the digital design, programmable latches were used to store synapse weights and full adders were employed to implement neuron [31]. For the analog design, memristive synapses were employed. Resistance values in the range of 10kΩ to 200kΩ were used to emulate memristors. In this design analog current-summers were employed for modeling the neuron [4]. The area was estimated based on the cross bar architecture for memristive neural network [4, 5].

TABLE 1
Design Performance for Character Recognition

| $N_n = 86$ $F_s = 500MHz$ | CMOS 45nm (digital) | CMOS 45nm (analog) $TiO_2$ memrstive synapse | Spin-CMOS Hybrid ANN |
|---|---|---|---|
| Power | 38mW | 63mW | 2.2mW |
| Network Area | 2.1 mm$^2$ | 0.010 mm$^2$ | 0.018 mm$^2$ |

TABLE 2.a
Spin ANN Specs

| Spin based ANN | |
|---|---|
| Number of Neurons* | |
| Input layer | 24 |
| Hidden Layer | 24 |
| O/P layer | 36 |
| Supply Voltage | |
| $Vdd_H$ (mV) | 825 |
| $Vdd_L$ (mV) | 800 |
| Current per firing operation per synapse | 10µA |
| # Weight Levels | 16 |

TABLE 2.b
CMOS ANN Specs

| Digital ANN | |
|---|---|
| #Full Adders | 4400 |
| Programmable Latches | 6240 |
| weight bit | 3 + 1 (sign) |
| Analog ANN | |
| Synapse resistance | 200K Ω 10 KΩ |
| Max. synapse Current | 4µA |
| Avg.Neuron power | 0.45mW |

TABLE 4
Device Parameters

| $K_{u2}$(biaxial anisotropy) | $2 \times 10^6$ erg/cm$^3$ | DWM polarization constant | 0.9 |
|---|---|---|---|
| Magnet size (nm$^3$) Neuron | 40x30x1 | Damping coefficient | 0.007 |
| Magnet size (nm$^3$) DWM | 350x80x10 | Channel material | Cu |
| $H_k$(coercivity) | 5KOe | Spin Flip Length | 1µm (300K) |
| $M_s$(saturation magnetization) | 400emu/cm$^3$ | resistivity | 7Ω-nm |

Table-I compares the two designs with the proposed spin based neural network. The digital implementation consumes large area as well as power due to bulky neuron and synapse units. Note that, a fully parallel implementation for the digital ANN was chosen for the purpose of comparison. Area for the digital design can be reduced through sequential processing using smaller number of neuron units, but power consumption is expected to remain almost constant for a given throughput. The analog implementation with memristive synapse turns out to be the most inefficient in terms of power. However, it achieves a large improvement in area as compared to the digital design due to compact synapses and cross-bar architecture [4, 5].

The spin-CMOS hybrid implementation achieves both, low power as well as small area, comparable to that of the analog ANN. The power and area benefits of the proposed design can be ascribed to simple and compact spin devices that operate at ultra low supply voltages and mimic the neuron operation. Both, low energy consumption, as well as compactness is conducive to integration of large number of neurons for programmable computational networks for cognitive and Boolean computation. Table-2 provides some relevant design details. Finally table 3 enlists some of the critical device parameters used in the simulation.

## VII. SUMMARY

Spin device phenomena like, majority evaluation, hard-axis switching, and adjustable spin polarization strength of domain wall magnets, clubbed with appropriate clocking scheme can lead to an energy efficient model for neuron-synapse unit. The localized, ultra low voltage operation of neuron-synapse units, assisted with efficient circuit and architecture level design strategies for inter-neuron signaling and power gating can facilitate high degree of integration. The proposed spin-CMOS hybrid ANN design can be suitable for low power, programmable computation architecture for cognitive as well as Boolean applications.

### Acknowledgement

This research was funded in part by Nano Research Initiative and by the INDEX center.